\documentclass[namedreferences,optionalrh]{spr-sola}
\usepackage{graphicx}        
\usepackage[dvipsnames]{xcolor} 
\usepackage{color}           

\usepackage{xurl}
\usepackage{siunitx}
\usepackage{blindtext}
\usepackage{setspace}

\usepackage[
  breaklinks = true,
   colorlinks = true,
  urlcolor   = blue,
  citecolor  = blue,
  linkcolor  = red, ]{hyperref}
\usepackage{lmodern}  
  
\chardef\us=`\_
\newcommand{\kms}{km~s$^{-1}$}
\begin{document}
\begin{frontmatter}
\title{Counter-Streaming Velocities in a Quiescent Filament}

\author[addressref={aff1},corref,email={garimakarki31@gmail.com}]{\inits{G.}\fnm{Garima}~\lnm{Karki} \orcid{0009-0003-3193-7496}}
\author[addressref={aff2,aff3,aff4},corref]
{\inits{B.}\fnm{Brigitte}~\lnm{Schmieder} \orcid{0000-0003-3364-9183}}
\author[addressref={aff2},corref]
{\inits{B.}\fnm{Pascal}~\lnm{D\'emoulin} \orcid{0000-0001-8215-6532}} %
\author[addressref={aff1}]{\inits{P.}\fnm{Pooja}~\lnm{Devi} \orcid{0000-0003-0713-0329}}
\author[addressref={aff1},corref]{\inits{R.}\fnm{Ramesh}~\lnm{Chandra} \orcid{0000-0002-3518-5856}}
\author[addressref={aff5,aff6,aff7},corref]{\inits{R.}\fnm{Reetika}~\lnm{Joshi} \orcid{0000-0003-0020-5754}}
\address[id=aff1]{Department of Physics, DSB Campus, Kumaun University, Nainital -- 263001, India}
\address[id=aff2]{LIRA, Observatoire de Paris, Universit\'e PSL, CNRS, Sorbonne Universit\'e, Universit\'e Paris Cité, 5 place Jules Janssen, 92195 Meudon, France} 
\address[id=aff3]{LUNEX EMMESI Institut, SBIC, Kapteyn straat 1, Noordwijk2201 BB Netherlands
} 
\address[id=aff4]{SUPA, School of Physics \& Astronomy, University of Glasgow, Glasgow G12 8QQ, UK}
\address[id=aff5]{NASA Goddard Space Flight Center, Heliophysics Science Division, Greenbelt, MD 20771, USA}
\address[id=aff6]{Department of Physics and Astronomy, George Mason University, Fairfax, VA 22030, USA}      
\address[id=aff7]{Rosseland Centre for Solar Physics, University of Oslo, P.O. Box 1029 Blindern, N-0315 Oslo, Norway}
\runningauthor{Karki et al.}
\runningtitle{Counter-streaming velocities in a quiescent filament}
\begin{abstract}
Filaments$/$prominences are cold plasma ($\approx 10^{4}$ K) embedded in the solar corona, two orders of magnitude hotter. 
Filament plasma is structured by the magnetic field in thin elongated threads.  Counter-streaming flows have been observed. The aim of this paper is to characterize these flows.  
For that, we use high spatial resolution observations of spectral data obtained with THEMIS in H$\alpha$ and with IRIS in Mg II k lines on 29 September 2023.
We best detect counter-streaming flows in both the blue and red wings of these spectral lines. They are forming long Doppler shifted strands slightly inclined on the filament axis. The blue$/$red shift alternates across the strands at the arc second scale.
H$\alpha$ spectral profiles with large widths are interpreted as formed by multi-strands with opposite velocity directions. The absorption in the core of Mg II k line is also broader than in the chromosphere. This corresponds also to counter-streaming velocities.  We derive that a fraction of the filament plasma is moving at supersonic speed (of the order of 20~\kms) with the assumption that the filament is optically thick.
We conclude that the counter-directed Doppler shifts might not be magnetic field aligned flows but rather correspond to kink transverse oscillations of the magnetic field with independent motions in nearby strands.

\end{abstract}
\end{frontmatter}
 
\section{Introduction}
     \label{S-Introduction}

 Filaments, called prominences when observed at the limb, are made up of cold, partially ionized plasma embedded in a magnetic field that opposes gravity \citep{Tandberg-Hansen1974}. The formation and stability of filaments is still an active research field, and many physical questions remain.

\subsection{Filament Formation}
\label{subS:Filament_formation}

The consensus on filament stability is based on the key role of the magnetic field acting against gravity, with plasma frozen in the magnetic field. 
\citep[see reviews by][]{Mackay2010,Labrosse2010}. 
Various magnetic supports for cold plasma are proposed, such as sheared arcades \citep{Aulanier2002_DeVore}, in magnetic dips of quadrupolar magnetic configurations \citep{Kippenhahn1957,Demoulin1993} or in twisted flux ropes \citep{Kuperus1974,Priest1989,Aulanier1998_D,Aulanier1999}.
Magnetic flux ropes are often identified at the location of solar filaments by force-free nonlinear extrapolations \citep{Guo2010,Yan2016} or using magneto-hydrodynamic simulations
\citep{2016NatCo...711522J,Guo2024,Schmieder2024}. 
Finally, two different magnetic configurations can co-exist along the filament channel such as in the filament studied by \cite{Guo2010} or as the magnetic arcades of the inter-winding filament mixed with a flux rope proposed by \cite{Yan2025_intermediate}.

As for the formation of cold plasma, the debate continues over possible mechanisms, such as plasma condensation inside loops, levitation of cool plasma in rising magnetic field lines. There are many 2D and now 3D models of filament formation. Filaments form either by levitation of material from the photosphere \citep{Rust1994,Deng2000}, by condensation of coronal material in loops  \citep{Luna2012}, 
or by thermal instability \citep[e.g.][]{Xia2014,2022NatAs...6..942J,Donne2024}.

\subsection{Dynamics of the Plasma}
\label{subS:Dynamics_plasma}

Different methods exist to derive the velocities of the plasma in the filament. A first method is using time-slice to estimate plasma motion along the threads or in the plane of the disk in prominences. This is a classical method, especially with the high cadence images available nowadays. This method involves placing a virtual slit along the expected path of the plasma motion in a sequence of solar images and stacking the intensity profiles along this slit over time to create a two-dimensional time–distance (or time–slice) diagram.
However, the estimated velocity can be due to waves and not to plasma motion. Recently, observing a prominence at the limb with the high spatial resolution of the New Vacuum Solar Telescope (NVST) counter-streaming flows between 4 and 11 km s$^{-1}$ were measured with the time-slice method \citep{Yan2025}.

A second method is the bisector method applied to spectral line profiles \citep{Schmieder1991}. This method determines the global wavelength shift of spectral lines by constructing its bisector. The bisector is the wavelength at the midpoint between the red and blue wings at a given depth of the spectral line profile. The displacement of the bisector from its rest position (zero velocity), provides an estimation of the velocity. The derived Doppler shifts are typically minimum values due to partial cancellation of blue and red shifts and to the contribution of the chromospheric background in the line formation in the case where the filament is not optically thick.
 
A third method derives the line-of-sight (LOS) velocity by using a ‘cloud model’ technique \citep{Beckers1964,Schmieder1984,Mein1988,Kuckein2016}. This method requires the determination of the background profile with good accuracy.  \cite{Schmieder1991} and \cite{Schmieder2004} applied this method to the observed full profile of the H$\alpha$ line in filaments and found higher values  by a factor of 3 to 5 than those derived by the bisector method. This is how counter-streaming flows of the order of 15 \kms\ were measured in the feet of a  filament using the cloud model technique  and interpreted as part of the injection model \citep{Schmieder1991},  while  in the observations of \citet{Schmieder2004} they concerned the activity of two filament segments merging together.

A fourth method is to study maps taken at different wavelengths in the H$\alpha$ profiles. They are called Doppler images.
\cite{Deng2002} showed that Doppler images have generally higher contrast and finer structural details than the corresponding images in H$\alpha$ intensity. The structures in the Doppler images outline both filament ‘spines’ and the branching threads. The filament channel in these Doppler images is generally larger than the H$\alpha$ filament. This channel is formed of aligned threads, alternating blue and red shifted \citep{Aulanier2002}. The measured amplitudes of the LOS motions are generally less than 2 km s$^{-1}$, except during the periods of filament activation. The maximum detected blue shifts were around 4 km s$^{-1}$ and the maximum red shifts around 7 km s$^{-1}$  \citep{Deng2002}. 

In this paper we utilized both time-slice method for AIA data to derive transverse velocities, and maps at different wavelengths in the H$\alpha$ and Mg II k profiles to derive Doppler velocities.

\subsection{Counter-streaming Velocities}
\label{subS:Counter-streaming}

In order to analyze the counter-streaming flows in filaments, it is important to have the context of the observations, mainly the activity phase of the filament and also the spatial resolution of the instruments.  In  this paper we consider only quiescent filaments and prominences.  The spatial resolution is crucial to derive a conclusion on the counter-streaming flows.

The term ``counter-streaming'' was used for the first time  for a filament observed at Big Bear Solar Observatory in the paper of \cite{Zirker1998_c}. They found counter-streaming in the images obtained by a filter in a narrow band of wavelengths  in  the H$\alpha$ center and  at $\pm$ 0.3 \AA\ using what we call above method four.  
Their observations concerned a filament observed on the disk; therefore, they have to assume that the optical thickness of the filament was large enough to avoid the contamination of the chromosphere below. Nevertheless, they found adjacent vertical sheet/threads with up and down flows with a speed close to a few \kms. They conclude by  this sentence: $``$as the flows must be aligned with the magnetic field, this observation implies that a field connects the prominence directly to the photosphere".  This conclusion is contradictory to the common view of the horizontality of the magnetic field lines sustaining filament plasma in dips \citep{Aulanier1998_BP,Aulanier1999}. A similar work using maps in the H$\alpha$ wings and time-slice method was next achieved by \cite{Deng2002}.

Later on, \cite{Schmieder2010} detected counter-streaming in a prominence observed by the MSDP  operating at the solar tower in Meudon; they were identified in H$\alpha$ Doppler maps computed with a bisector method.  The prominence was also observed by Hinode$/$SOT, which allowed them to compute the velocity of  blobs in the plane of sky perpendicularly to the LOS. The advantage of a prominence is that there is no  contamination from the chromosphere.
The calculated  Doppler shifts  and the blob velocities were of the order of 3 \kms\   similarly to those of \cite{Zirker1998_c}. 
The reconstructed  velocity vectors showed a high inclination on the prominence threads. %
The Doppler shift areas in the prominence were wider (close to 10$''$)  than the thread intensity cross-section,  meaning that the plasma motions inside adjacent threads were coherent in a larger volume than in an individual thread. 
This conclusion may be due to the difference in resolution of Hinode  (0.5$''$) and the MSDP (1$''$).  

Next, a few studies tried to explain the presence of counter-streaming in magnetic field dips in a prominence.  \citet{Shen2015} proposed that the counter-streaming flows in the horizontal magnetic field lines of prominence barbs were
caused by the imbalanced pressure at both ends.
In contrast, \citet{Zhou2020} proposed that counter-streaming is due to turbulent heating in the low solar atmosphere.  With such a setup, they performed a numerical simulation showing random evaporation and formation of sparse threads in prominences. Previously it has been demonstrated that in  thermal non-equilibrium models, condensations form  along long, low-lying magnetic field lines  and develop  transient high-speed motions and elongated threads \citep{Antiochos2000,Mackay2001,Karpen2006}. These models of evaporation -condensation could also explain counter-streaming flows in horizontal threads. Jet interaction with filament also develops oscillations along the filament threads as shown in \citet{Nobrega2024}.

\begin{figure}    
\centerline{\includegraphics[width=1.\textwidth,clip=]{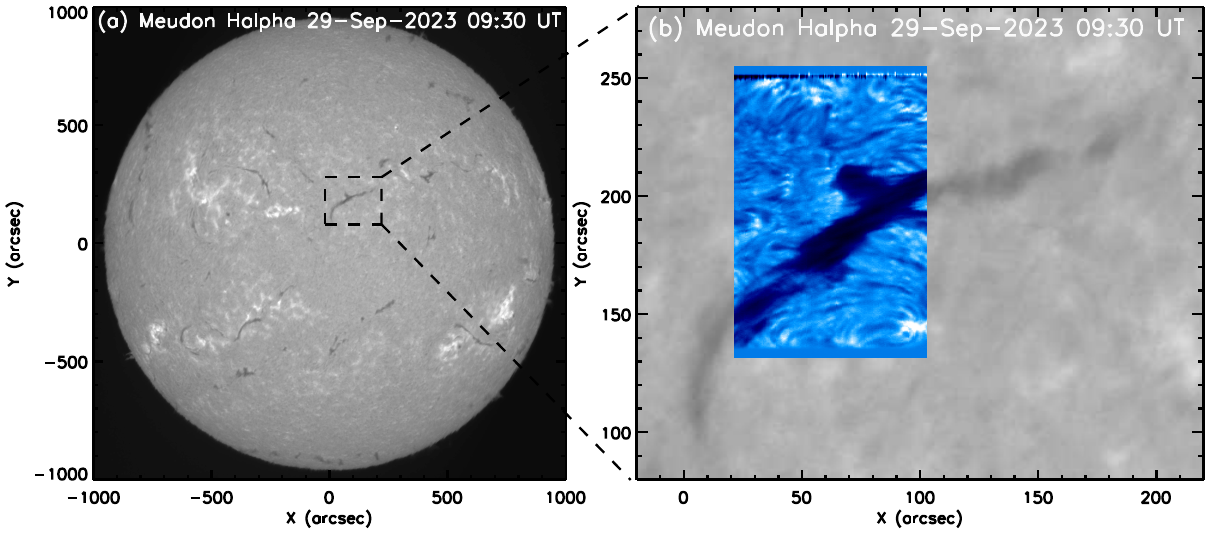}}
       \caption{Panel (a) shows the full disk H$\alpha$ image observed by Meudon spectroheliograph on 29 September 2023. The zoomed view of the dashed black box is shown in panel (b), along with the co-aligned filament observed by THEMIS in blue. 
       }
\label{Meudon}
\end{figure}

We used H$\alpha$ observations obtained with the T\'elescope H\'eliographique pour l’Etude du Magn\'etisme et des Instabilit\'es Solaires  (THEMIS new generation; \citet{Schmieder2025}) and its adaptive optics (pixel size = 0.06$''$) joined with space data in Mg II k lines with the Interface Region Imaging Spectrograph (IRIS). It is the first time that we have simultaneously high spatial  and spectral resolution observations of the  fine structures in a filament.

\subsection{Paper Outline}
\label{subS:outline}
  
The paper is organized as follows. In Section \ref{Overview} we present the characteristics of the instruments used and the outlines of the studied observations. In Section \ref{Spectroscopy} we analyze 2D maps at different wavelengths in the H$\alpha$ and Mg II k lines, and associated high-resolution spectra. These data are complemented with AIA and HMI observations.  The various results are interpreted within the framework of the typical magnetic configuration of filaments. Finally, in Section \ref{Conclusion} we summarize our findings and discuss the coherence of the various results.

\section{Overview of the Observations}
\label{Overview}
For the present study of a filament observed on 29 September 2023, we have used high-resolution data from the following space- and ground-based observatories.
The Solar Dynamics Observatory \citep[SDO;][]{Pesnell2012} includes two key instruments: the Atmospheric Imaging Assembly \citep[AIA;][]{Lemen2012} and the Helioseismic and Magnetic Imager \citep[HMI;][]{Schou2012}. AIA captures full-disk images of the Sun in seven extreme ultraviolet (EUV), two ultraviolet (UV), and one white-light wavelengths, with a pixel size of 0.6$''$ and a temporal resolution of 12 s. For this study, we specifically utilized AIA 193 \AA\ data to analyze the filament. HMI, on the other hand, provides observations of the Sun at 6173 \AA\ with a pixel size of 0.5$''$ and a temporal resolution of 45 s. In this study, HMI magnetograms were used to examine the magnetic configuration of the filament region.

The Interface Region Imaging Spectrograph \citep[IRIS;][]{DePontieu2014} observes the interface region between the relatively cool solar surface (photosphere, temperature $\approx$ 6000 K) and the extremely hot outer atmosphere (corona, temperature $\approx 10^6$ K). IRIS captures slit-jaw images in four UV wavebands: Mg II wing (2830 \AA), Mg II k (2796 \AA), C II (1330 \AA), and Si IV (1400 \AA). Additionally, it records spectra within pass-bands of 1332–1358 \AA, 1389–1407 \AA, and 2783–2834 \AA, which includes prominent spectral lines such as Mg II h (2803 \AA) and Mg II k (2796 \AA) formed in the chromosphere, as well as C II (1334/1335 \AA) and Si IV (1394/1403 \AA) formed in the transition region. IRIS can simultaneously capture images and perform spectral rasters, sampling regions up to 130$''$ × 175$''$ with various spatial resolutions. The mode of observation of  IRIS during our filament campaign consists of slit-jaw images with the Mg II filter and dense coarse rasters of 64 steps. IRIS was centered at x = 74$''$ and y = 175$''$ and obtained three rasters from 10:34 UT to 10:51 UT, from 10:51 UT to 11:08 UT,  from 11:08 UT  to  11:26 UT. The raster step size is 2$''$ so that each spectral raster spans a field-of-view (FOV) of $127'' \times 175''$ in $\approx$ 17 minutes.

\begin{figure}    
\centerline{\includegraphics[width=1.\textwidth,clip=]{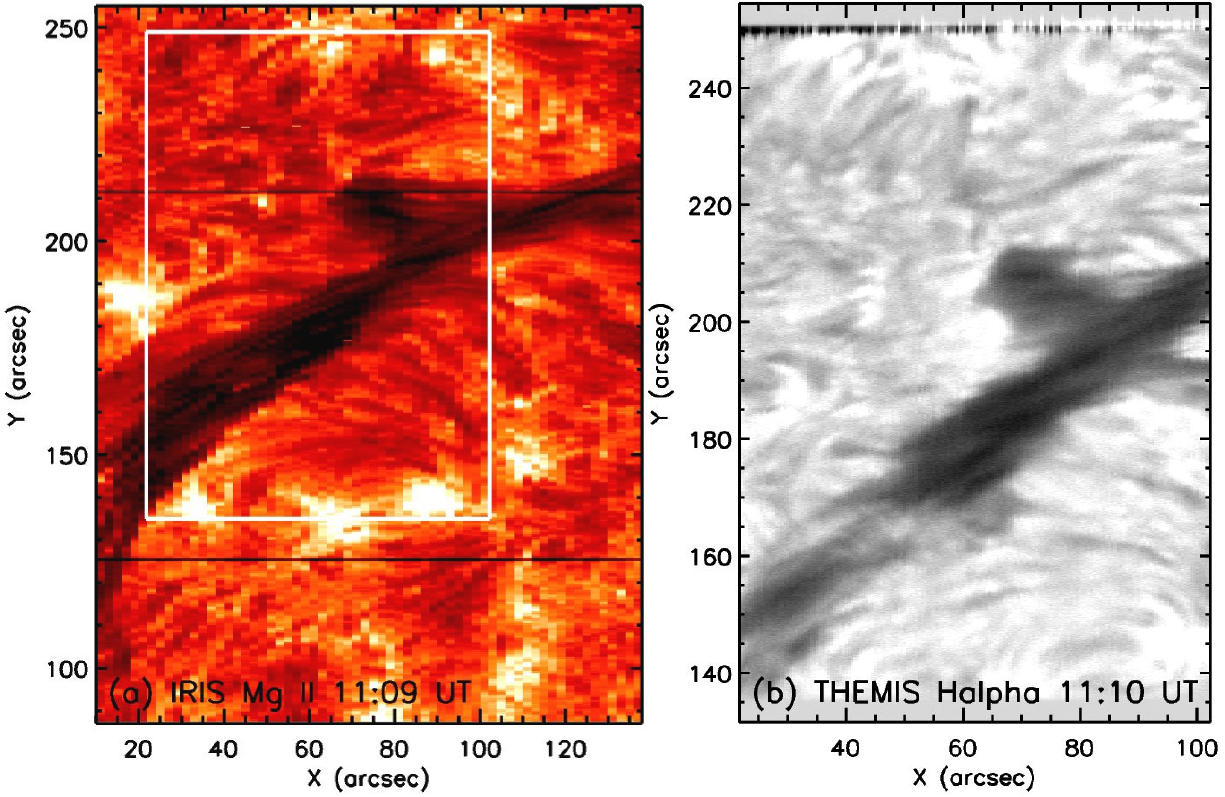}}
\small
       \caption{Filament observed by IRIS in Mg II k line (panel a), and  by THEMIS in H$\alpha$ around 11:10 UT (panel b). These images have been reconstructed from rasters. The white box in panel (a) shows the FOV of THEMIS of panel (b). In the two maps the filament has internal strands which are making a small angle with the filament axis. Their direction contrasts, especially in IRIS map, with the fibril direction seen outside of the filament and with coherent directions on both sides.}
\label{THEMIS_IRIS}
\end{figure}

\begin{figure}    
\centerline{\includegraphics[width=0.9\textwidth,clip=]{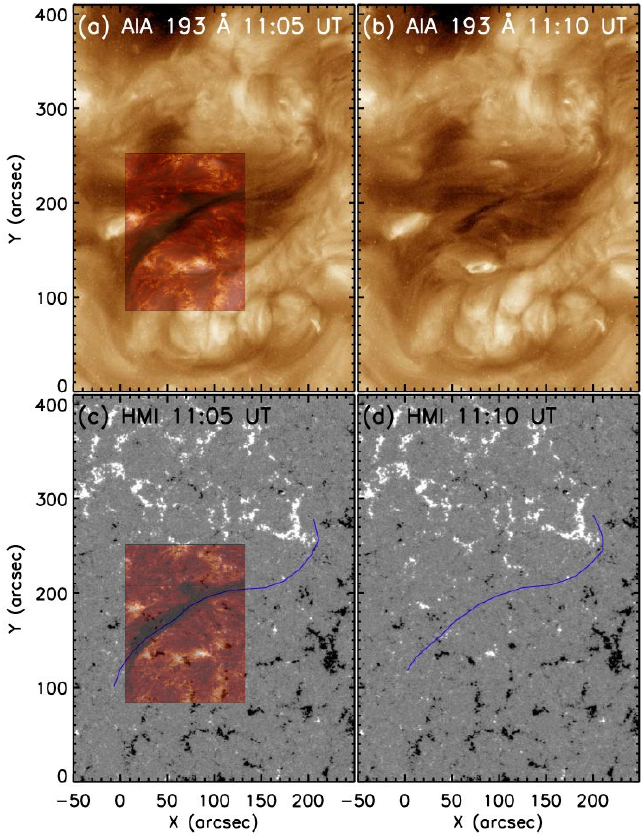}
}
\caption{AIA 193 \AA\ images (panel a, b), HMI magnetogram (panel c, d). The IRIS raster image is overlaid on AIA 193 \AA\ and HMI in panels (a) and (c). 
The bright points at the bottom of IRIS images are over the boundary of the magnetic network and over the bright points in AIA 193 \AA. An animation of the AIA data is available.}
\label{AIA_HMI}
\end{figure}

\begin{figure}    
\centerline{\includegraphics[width=1.\textwidth,clip=]{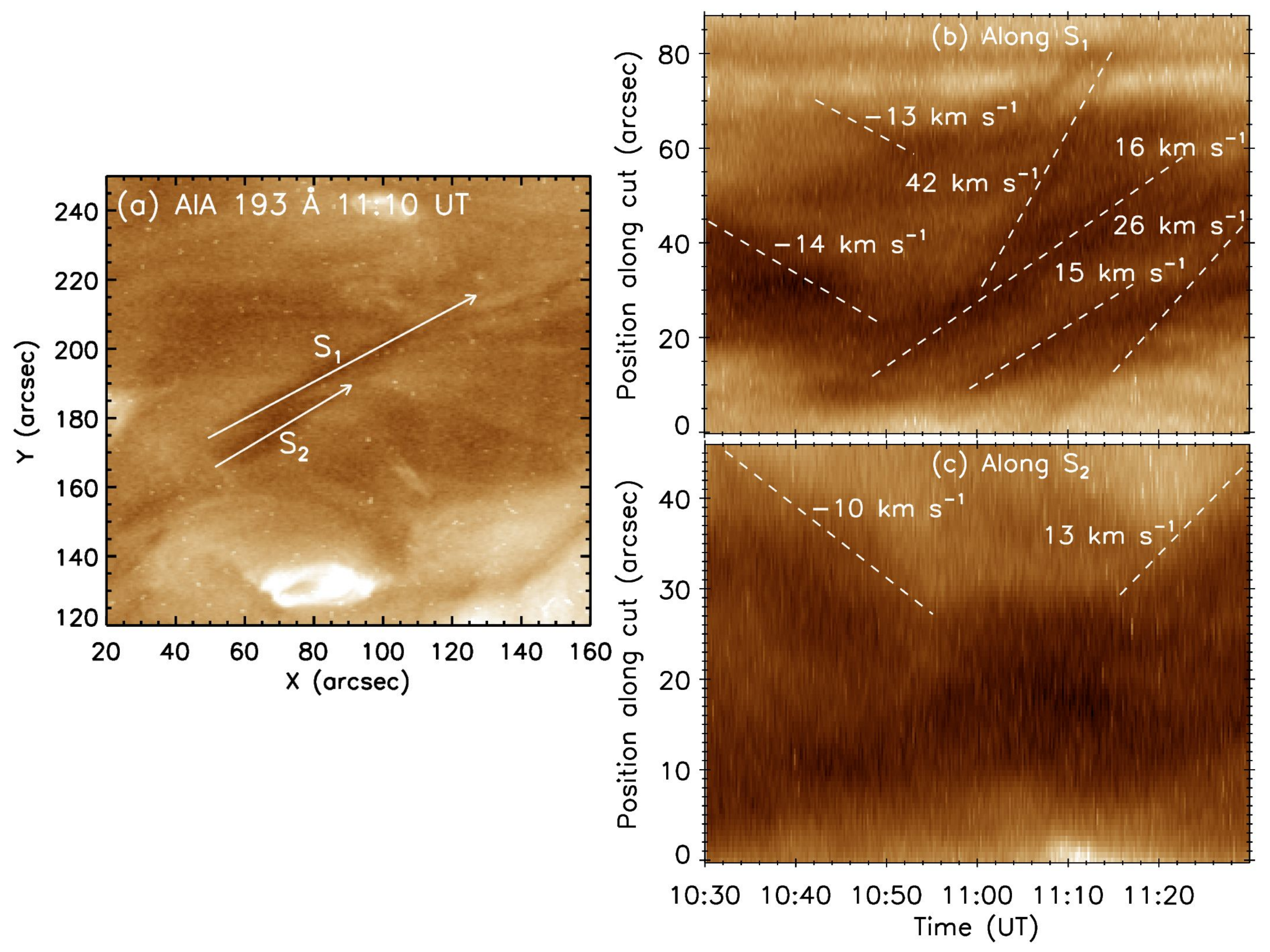}}
\caption{Time-slice diagram along the sigmoid in the filament channel observed in the AIA 193 \AA\ filter. Panel (a) shows the positions of the slices S$_1$ and S$_2$. The time-distance diagrams corresponding to S$_1$ and S$_2$ are shown in panels (b) and (c), respectively. 
Several moving structures are identified with dashed lines and their plane of sky speeds are included.}
\label{time_slice}
\end{figure}

The T\'elescope H\'eliographique pour l’Etude du Magn\'etisme et des Instabilit\'es Solaires \citep[THEMIS;][]{Mein1985} is installed in the Canary Islands. The primary mirror is 92 cm diameter.  The telescope is equipped with highly flexible spectroscopic capabilities that enable simultaneous observations of several spectral ranges. Here we used only the spectrum centered on the H$\alpha$ line with a CMOS camera (2k $\times$ 2k). Details of mode of observations can be found in \citet{Karki2025}. 
The pixel resolution along the slit is 0.06$''$. 
The slit width is 0.5$''$ and the stepping is either 0.5$''$  or 1$''$. The spectral dispersion is $\sim$ 3.076 m\AA\ per pixel. The bandpass is 6.3~\AA\ around 6563~\AA. The exposure time for H$\alpha$  varies from 0.05 to 0.2 s depending on targets and seeing conditions. The  adaptive optics  \citep{Tallon2022} are used, which allows us to have high-spatial resolution data \citep{Schmieder2025}.

Meudon spectroheliograph provides full disk images of the Sun in H$\alpha$ and Ca II wavelengths derived from line profiles with pixel size and temporal resolution of 1.1$''$ and 1 minute, respectively. For the current study we have used H$\alpha$ spectroheliograph to study the filament.

The target of IRIS was a small filament near the disk center. THEMIS followed this target between 10:59 UT and 11:12 UT with a cadence of 2 minutes, obtaining  eight rasters of 160 spectra each  with a step of 0.5$''$ and an exposure time of 0.2~s  allowing comparison with  space-borne observations (IRIS and SDO) and  Meudon spectroheliograph (Figure \ref{Meudon}).
The co-alignment with IRIS was achieved by aligning the slit-jaw images in Mg II k with AIA 304 \AA. Specifically, the bright points observed in AIA 304 \AA\ are matched with those in the slit-jaw images. THEMIS is then co-aligned with IRIS by matching the filament contours in both datasets. As a result, THEMIS is aligned with both IRIS and AIA. The co-alignment with IRIS allows us to put coordinates in the THEMIS FOV (80$'' \times 120''$, Figure \ref{THEMIS_IRIS}). 
The alignment of IRIS/THEMIS images at different wavelengths is automatic.
Finally, in this study, we focus mainly on comparing the strand orientations; therefore, a shift of 5$''$ (for example) between the IRIS and THEMIS datasets would not affect the conclusions.

Within IRIS Mg II  and in THEMIS H$\alpha$ maps, the filament exhibits internal strands which make an angle of around 20 degrees with the filament axis, value which is in good agreement with previous studies \citep[e.g.][]{Leroy1984, Aulanier2003}. Their direction contrasts with the fibril direction seen in the nearby chromosphere, with coherent directions on both sides of the filament.
From that we conclude that the optical depth $\tau$ is larger than 1 in the filament (so we do not see the chromosphere below the filament). This is important to detect the velocity in the filament strands since they are not altered by the dynamics of the chromosphere. 

Figure \ref{AIA_HMI} shows the overlay of the IRIS filament on AIA 193 \AA\ images and HMI magnetograms. 
The co-alignment is good since the Mg II brightenings, on both sides of the filament,  overlie magnetic polarities of the network.
The filament lies between positive and negative magnetic polarities. The barbs and the shape of the reversed S sigmoid indicate a dextral filament, with negative magnetic helicity according to the definition found in  \citet{Martin1998,lopez2006}. The sigmoid is the trace of the thicker part of the filament because the optical thickness of the absorption in 193 \AA\ is equivalent to the optical thickness  in H$\alpha$ \citep{Anzer2008,Heinzel2008}. From previous studies, the dense plasma of the filament is expected to be located in the magnetic dips of a flux rope \citep{Aulanier1998_D,Aulanier1999}. Indeed, the sigmoid location is similar to the one of a flux rope embedded in a bipolar magnetic region as found in MHD simulations \citep{Aulanier2010,Donne2024}.  Then, this filament is expected to be located within a flux rope anchored in the north-west in a hook ending in positive magnetic polarities and going to the south-east towards dispersed negative polarities.  

The studied filament was already visible two days before and is still visible the next day on 30 September. Then, it is relatively stable and it consists of several sections which are changing slowly of shape, in a time scale of an hour (strands) to a day (global shape).   This is shown in the Appendix (Figure \ref{IRIS_2}) with three IRIS raster maps obtained in the blue and red wings of Mg II k line during about half an hour. They show a slight evolution of the stands.  During our quarter hour  of H$\alpha$ observations, changes are marginal. Below, we mainly use the data obtained at 11:10 UT.
In  the 193 \AA\ movie some connections above the filament are present between bright areas on the left and right sides. Material is also flowing along the sigmoidal filament. With time-distance diagrams we measure  motions in both directions along the slices set along the filament (Figure \ref{time_slice}). The velocity of the blobs along the filament, projected in the plane of sky, is between about 10 and 40 \kms . 

\begin{figure}    
\centerline{\includegraphics[width=\textwidth,clip=]{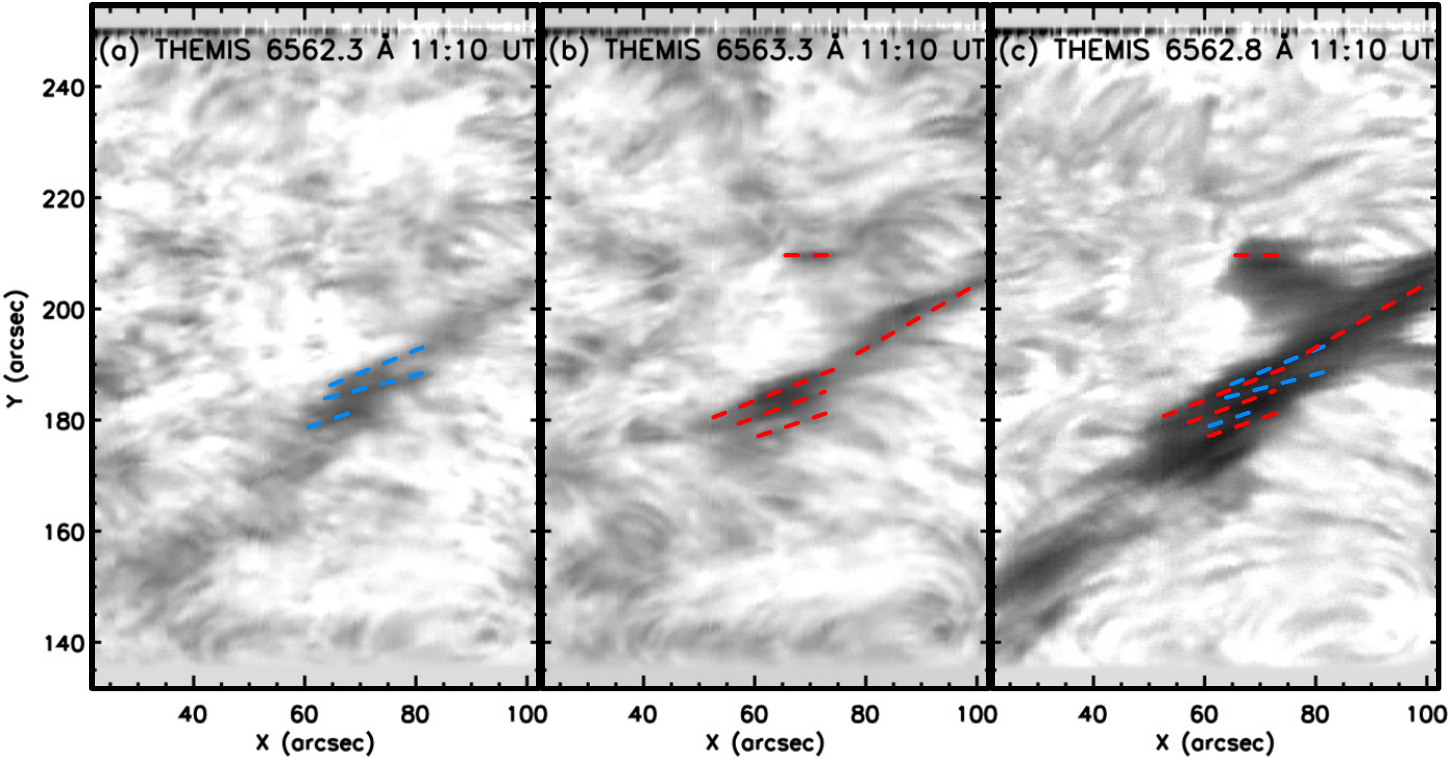}}
\caption{THEMIS H$\alpha$ maps at different wavelengths in H$\alpha$ at 11:10 UT in the line center and at $\pm$ 0.5 \AA\ which are the locations of the inflection points in the H$\alpha$ profile. These wavelengths correspond to (a) the blue wing at -22 \kms, (b) the  red wing at 22 \kms, (c) the line center. The blue and red dashed lines correspond to the thread-like structures or Doppler strands observed in the blue and red wings of H$\alpha$, respectively. Their positions are compared in panel (c).} 
\label{map_ha}
\end{figure}

\begin{figure}    
\centerline{\includegraphics[width=1\textwidth,clip=]{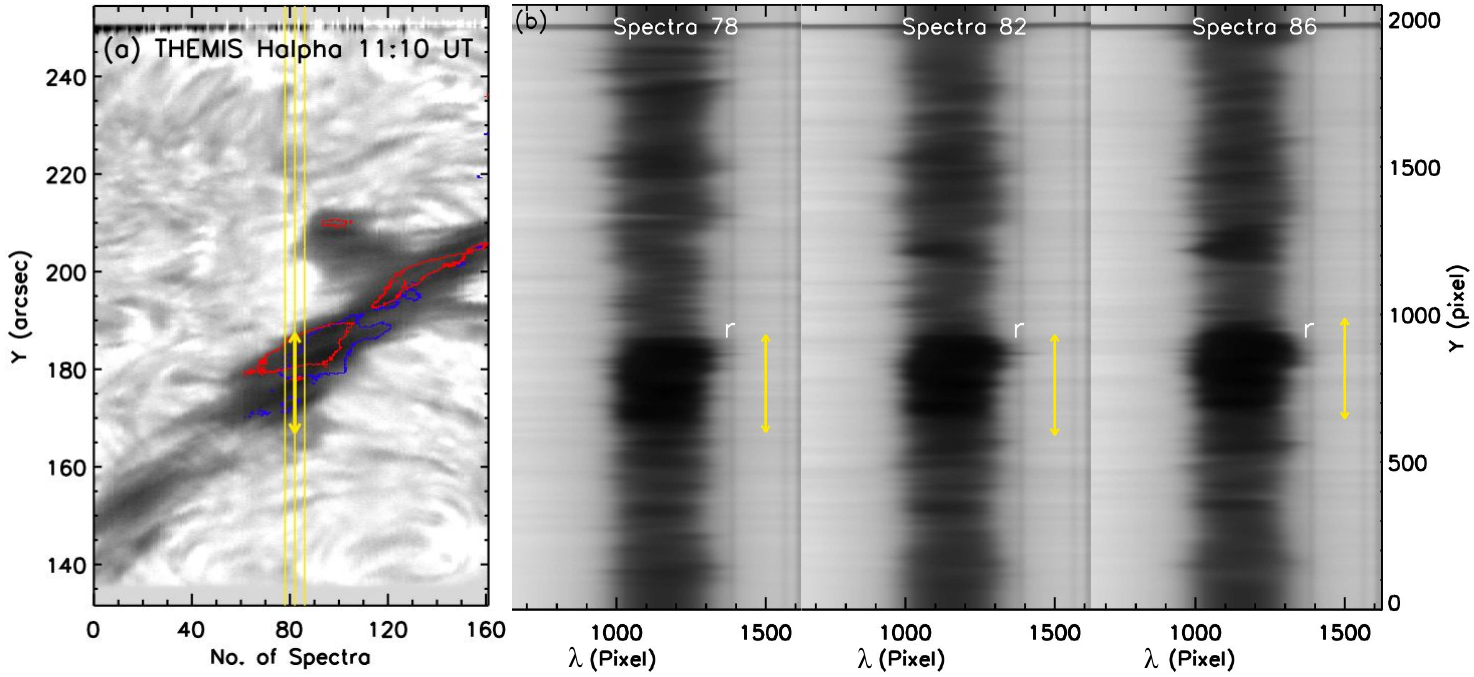} }
\caption{THEMIS H$\alpha$ map (panel a) and three filament spectra (panel  b). The blue/red contours in H$\alpha$ map  correspond to isocontours defined in the blue wing image at 6562.3 \AA\ and the red one at 6563.3 \AA\ respectively (these images are shown in Figure \ref{map_ha} a, b).
The yellow vertical lines in panel (a) shows the positions of the spectra (panel b).  The distance between successive positions is of 2$''$. The double headed arrow in all the panels indicates the cross-section of the filament for spectra 78, 82 and 86. These three spectra show the spatial continuity of red shifts, for example, along the filament Doppler strands (r position).
}
\label{spectra1}
\end{figure}

\section{Spectroscopy Results}
\label{Spectroscopy}

\subsection{H$\texorpdfstring{\alpha}{α}$ Maps at Different Wavelengths}  
\label{subS:Halpha_maps}

Figure \ref{map_ha} presents the filament observed in the H$\alpha$ line center (c) and in two wavelengths  at $\pm$ 0.5 \AA\ of the line center (a, b), which correspond to the wavelengths at the inflection point in the H$\alpha$  profile and to Doppler shifts of $\pm$ 22 \kms.  This velocity is comparable to the sound speed, $\approx 15$ \kms, in a plasma at a temperature of $\approx 10^4$ K. In fact the Doppler velocity is only one velocity component so such flows are well supersonic.  
In the blue and red wing images parallel Doppler strands have been identified and overlaid by dashed short lines in blue and red. 
In panel (c) these lines are overlaid on the observation in H$\alpha$ line center. This shows their miss-match, which is an indication of counter-streaming. They are parallel to the intensity strands. 
Their length is around 10 to 20$''$. The distance between red strands is less than 5$''$ and the distance between blue and red is of the order of 1$''$.

In Figure \ref{spectra1} (panel a) intensity isocontours, taken in the blue and red H$\alpha$ wings, are overlaid on the H$\alpha$ line center image. These isocontours are obtained for DN= 39000 and 38000 values, respectively, at the inflection points of the H$\alpha$ reference profile (see Section~\ref{subS:Halpha_spectra}).
The two contours overlie the central part of the filament along the axis rather than the filament edges.  
Indeed, at the filament borders, the fine structures are more isolated, so that only either blue shifted or red shifted plasma is observed. The blue shifted area is mostly in the southern part, and red shift is in the northern part along the filament axis.
Finally, the velocity in the plane of sky and along the narrow Doppler strands cannot be computed with a time-slice method because the x step is 0.5$''$ and the cadence is 2 minutes. Thus the spatial resolution perpendicular to the slit is not sufficient to measure  the velocity along the strands.
 
The regions along the LOS between the filament and the observer have low optical depth, $\tau$; because their contribution is proportional to $\tau$, they make only a small contribution to the observed spectral profile. The regions with large $\tau$ values, typically the region along the LOS including the filament bottom part and below, are also negligibly contributing to the observed spectral profile since their contribution is proportional to e$^{-\tau}$ (their emission is absorbed by the upper layers). For the studied filament, $\tau$ at the bottom of the filament is large since we do not see traces of chromospheric structures in the filament. Thus, we make the assumption that the observed profiles give information on the plasma with $\tau \approx 1$, which is located at some height in the filament. A deeper analysis on another quiescent filament indeed shows that $\tau$ is typically between 1 and 2 \citep{Kuckein2016}.

\begin{figure}    
\centerline{\includegraphics[width=1\textwidth,clip=]
{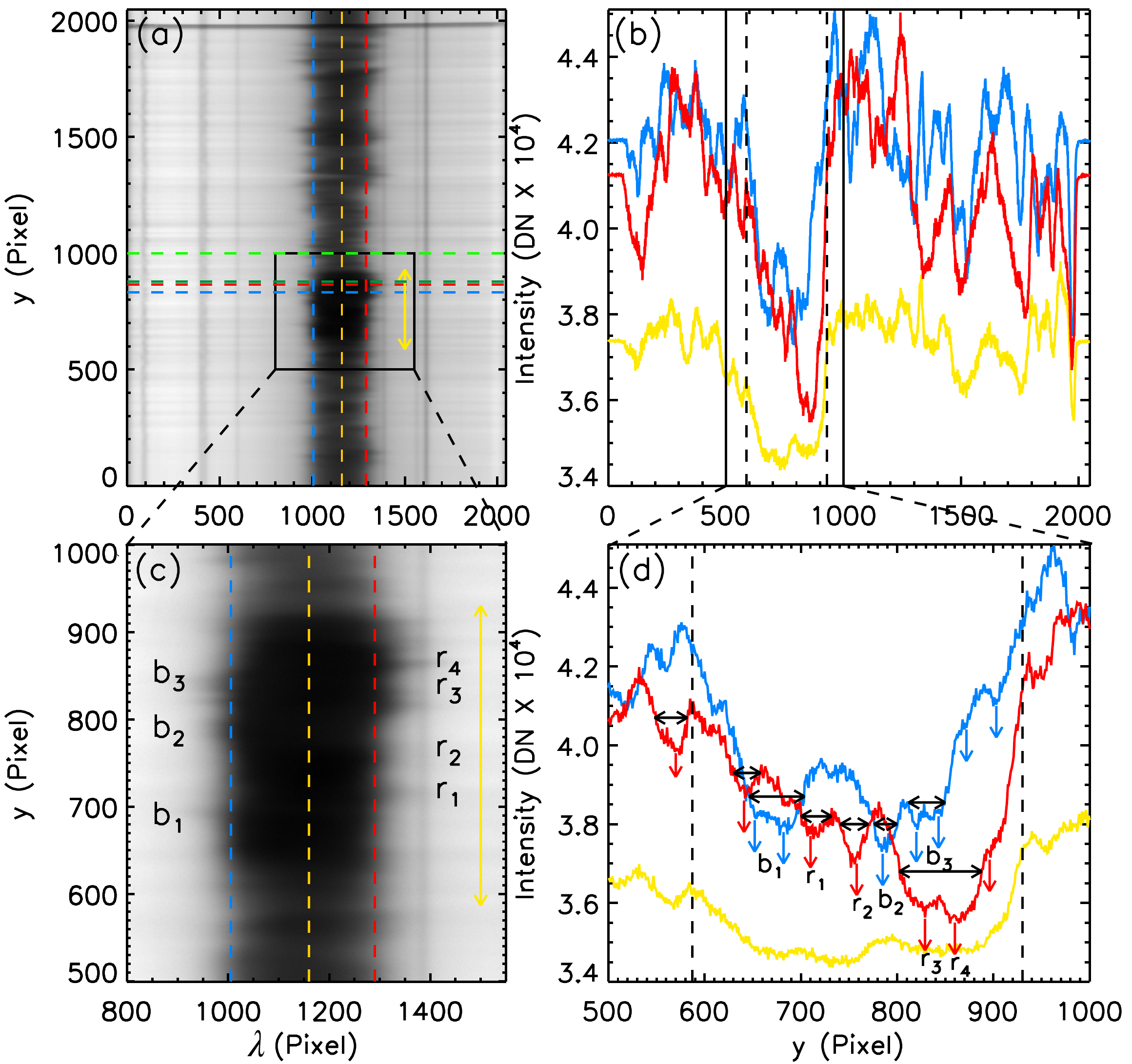}}

\caption{H$\alpha$ spectrum ($\lambda$, y) at 11:10 UT for the spectra  78 (panel a) and cuts along the slit at three wavelengths (panel b) showing the velocity shift in the filament and the chromosphere. The x location of spectra  78 is shown in Figure \ref{spectra1}.  
~In panel (a), the dashed vertical lines with blue, yellow, and red colors correspond to velocity of -20, 0, and 20 \kms, respectively, showing the cut locations used for the curves with same colors of panels (b, d).
In panels (b, d), the vertical  dashed lines are the borders of the filament cross-section. 
In panel (b) the vertical continuous lines show the y extension of panel (d). 
Panel (c) shows a zoom on the H$\alpha$ spectrum on the rectangular region shown in panel (a). 
A zoom of panel (b) is shown in panel (d). 
In panel (d) the high Doppler strands in the blue wing (b$_1$, b$_2$, b$_3$) and in the red wing (r$_1$, r$_2$, r$_3$, r$_4$) are indicated by double black horizontal arrows.  The vertical arrows mark their Doppler shifts with alternatively blue and red arrows.
Furthermore, in panel (a), the red and blue dashed horizontal lines indicate the locations at which the red and blue shifted profiles shown in Figure \ref{spectra3} a, c are drawn, respectively. The dark green dashed line corresponds to the red and blue shifted profiles shown in Figure \ref{spectra3} b. Finally the light green dashed line corresponds to the reference profile taken outside the filament and plotted in green in Figure \ref{spectra3}.}
\label{cut_zoom}
\end{figure}

\subsection{H$\texorpdfstring{\alpha}{α}$ Spectra and Profiles}
\label{subS:Halpha_spectra}

Figure \ref{spectra1} (panel b) presents three spectra crossing the filament. We note that the H$\alpha$ line profile is broadened and darker at the location of the filament (yellow arrow) than in the nearby chromosphere. The three spectra are selected at close positions to show the continuity of the velocity pattern in the blue and red wings, in agreement with isocontours extension along the filament (Figure \ref{spectra1} a). Due to good seeing conditions there are particularly high spectral and spatial resolution spectra along the slit, taking advantage of its pixel size of 0.06$''$.

We next analyze cuts at fixed wavelengths in the blue and red wings of H$\alpha$ and at the line center (see the vertical dashed lines in Figure \ref{cut_zoom} panels a and c).  The intensities in DN along these cuts are shown in panels b and d. The filament position corresponds to the minimum intensity in the yellow curve (H$\alpha$ line center). In the blue and red intensities, very thin fluctuations as a function of distance along the slit are present in the chromosphere and in the filament. They are more visible in the cross-section of the filament within the zoom plot of Figure \ref{cut_zoom} d. The filament width in the y-direction is about 340 pixels (20$''$).  The typical width of the fine velocity structures is around  25 pixels (1.5$''$). They are located inside intensity strands of width $\approx$ 100 pixels (6$''$). Blue and red shifted fine structures alternate. Indeed, in each spectrum crossing the filament and the barb we detect high velocity flows in opposite directions in smaller Doppler strands than the intensity threads. In the Appendix we present another example of spectra crossing both the filament and the barb (Figure   \ref{cut_1105_Barb}). We conclude that Figures \ref{cut_zoom} and \ref{cut_1105_Barb} confirm quantitatively what is  shown in the  images done for the blue, center and red wavelengths (Figure \ref{map_ha}).

\begin{figure}    
\centerline{\includegraphics[width=0.95\textwidth,clip=]{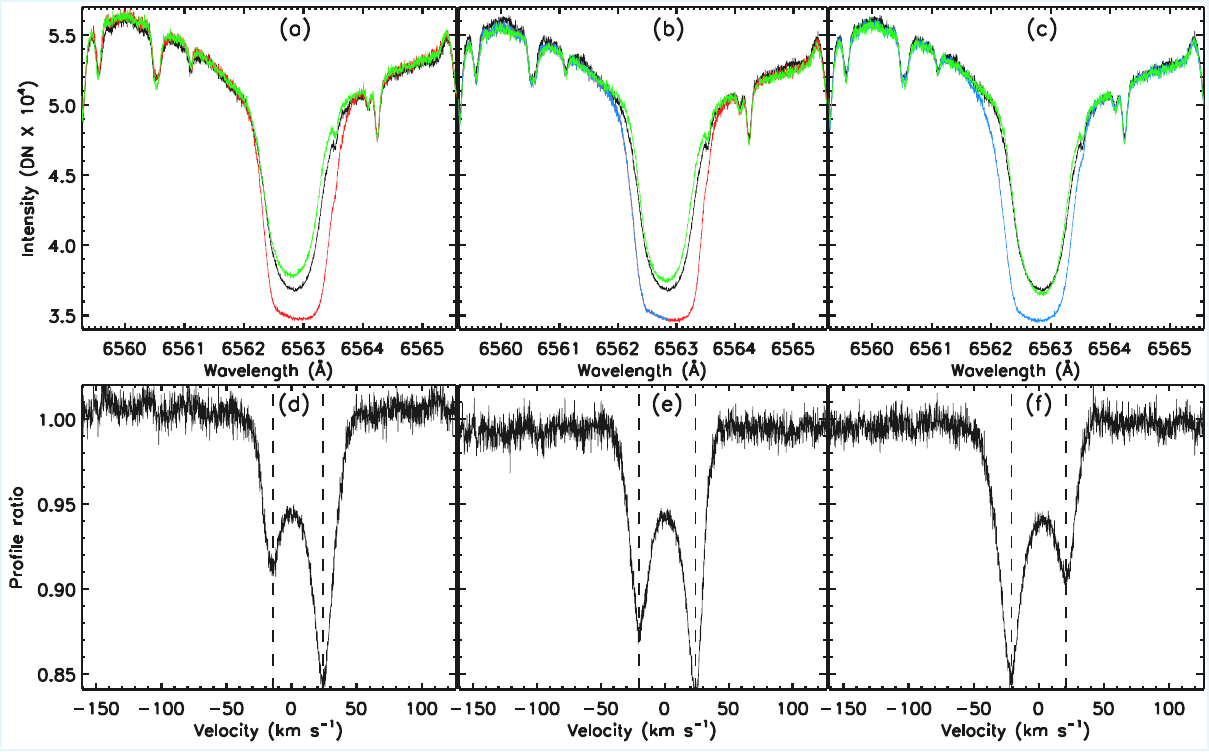}}
\caption{Examples of H$\alpha$ profiles in the filament (panels a, b, c) corresponding to spectra at spatial positions (x = 80, y = 865), (x = 86, y = 879) and (x = 95, y = 831) in pixel units, respectively. The corresponding x, y coordinates are shown in Figure \ref{cut_zoom}. 
The blue and red profiles are examples of blue and red shifted profiles in the filament, respectively. Their y positions are indicated  by red, blue and dark green dashed horizontal lines in panel (a) of Figure~\ref{cut_zoom}.
The reference profile in the chromosphere, in black, is selected from the mean spectra of a quiet region where the filament is not present. The light green profile is taken at pixel y=1000 in the chromosphere and is very similar to the reference profile. The inflection points in the reference profile are around $\pm$ 20 \kms. 
Panels (d--f) show the profile ratio (ratio with the reference profile) of the H$\alpha$ profiles of panels (a--c). 
Red$/$blue profiles show a greater absorption in the red/blue wing which corresponds to velocity between $\pm$~20 and $\pm$~30~\kms.}
\label{spectra3}
\end{figure}

\subsection{Interpretation of the H$\texorpdfstring{\alpha}{α}$ Profiles}
\label{subS:Interpretation}

Figure \ref{spectra3} gives examples of some broad H$\alpha$ profiles observed in the filament compared to  reference profiles obtained in the chromosphere outside the filament.
Two reference profiles are shown. The black profile is a mean spectrum of a quiet Sun region and the green profile is taken close to the filament in the chromosphere (see the light green dashed line in Figure \ref{cut_zoom}). 
Both reference profiles have narrow spectral profiles. The inflection point is around $\pm$ 0.5  \AA\ from the line center. In contrast, the spectral profiles in the filament have a very flat core indicating that they are likely formed by the sum of many profiles of threads having a large dispersion of velocities.  

The bottom row of Figure \ref{spectra3} shows the  ratio between filament profiles and the reference profile. They correspond to ratio profiles in the case of optically thick profiles. Indeed, chromospheric fibrils, seen on both filament sides, are not observed in the filament (Figures \ref{THEMIS_IRIS}, \ref{map_ha}). 
This implies that the spectra ratio provides information on the velocity distribution within the filament (where $\tau \approx 1$).  
We find Doppler shifts between 20 \kms\ and 30 \kms.  These are supersonic flows in a filament plasma at a temperature of $\approx 10^4$ K (the sound speed is $\approx 15$ \kms ).

\begin{figure}    
\centerline{\includegraphics[width=1\textwidth,clip=]{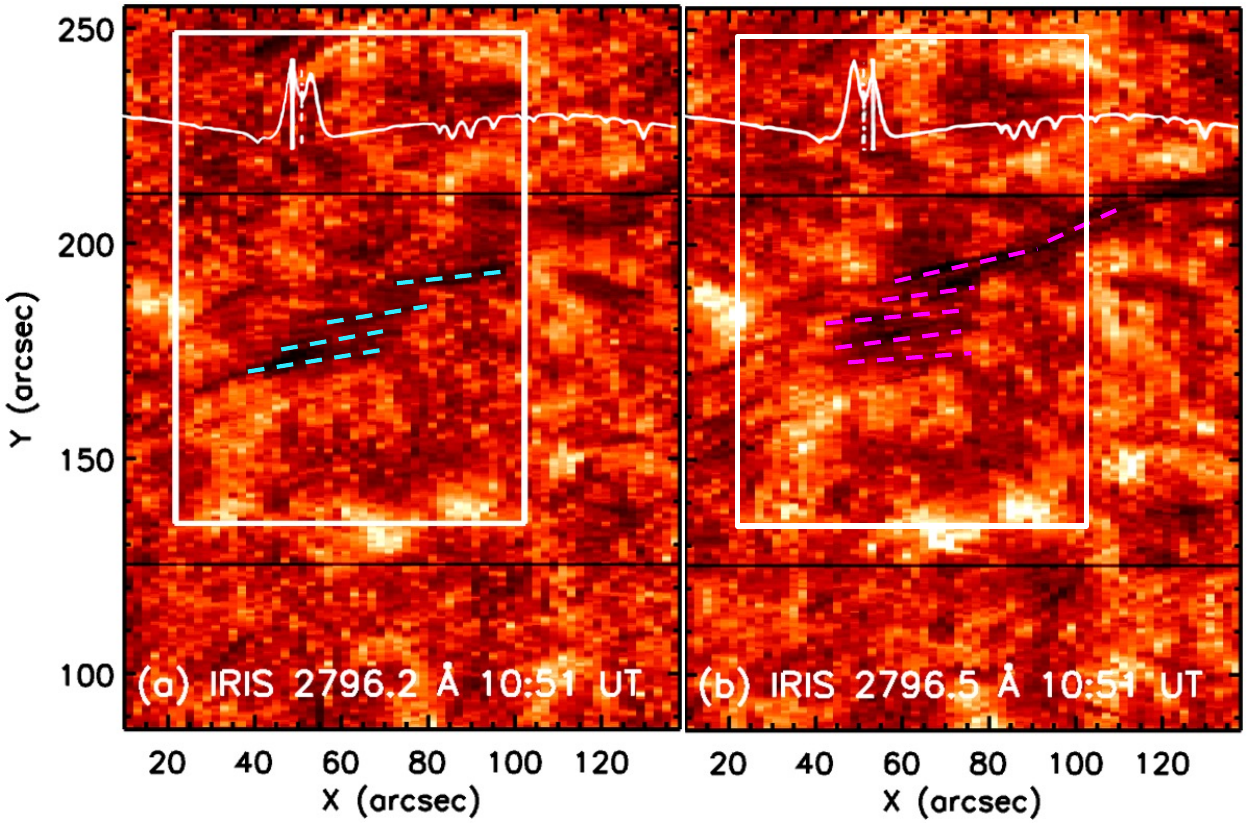}}
\caption{IRIS maps in the peaks of Mg II k line. 
The filament  is detected in both  peaks revealing  high speed components in the blue wing (left panel) and red wing (right panel) in parallel Doppler strands. The profile of Mg II line is drawn at the top of the images showing the concerned peak.  In each panel, dashed lines are overlaid to mark the most important Doppler strands. They are alternated  counter-streaming flows in the filament. The white box indicated the FOV of THEMIS (see Figure \ref{THEMIS_IRIS}).
}
\label{IRIS_maps}
\end{figure}

\begin{figure}   
\centerline{\includegraphics[width=1\textwidth,clip=]{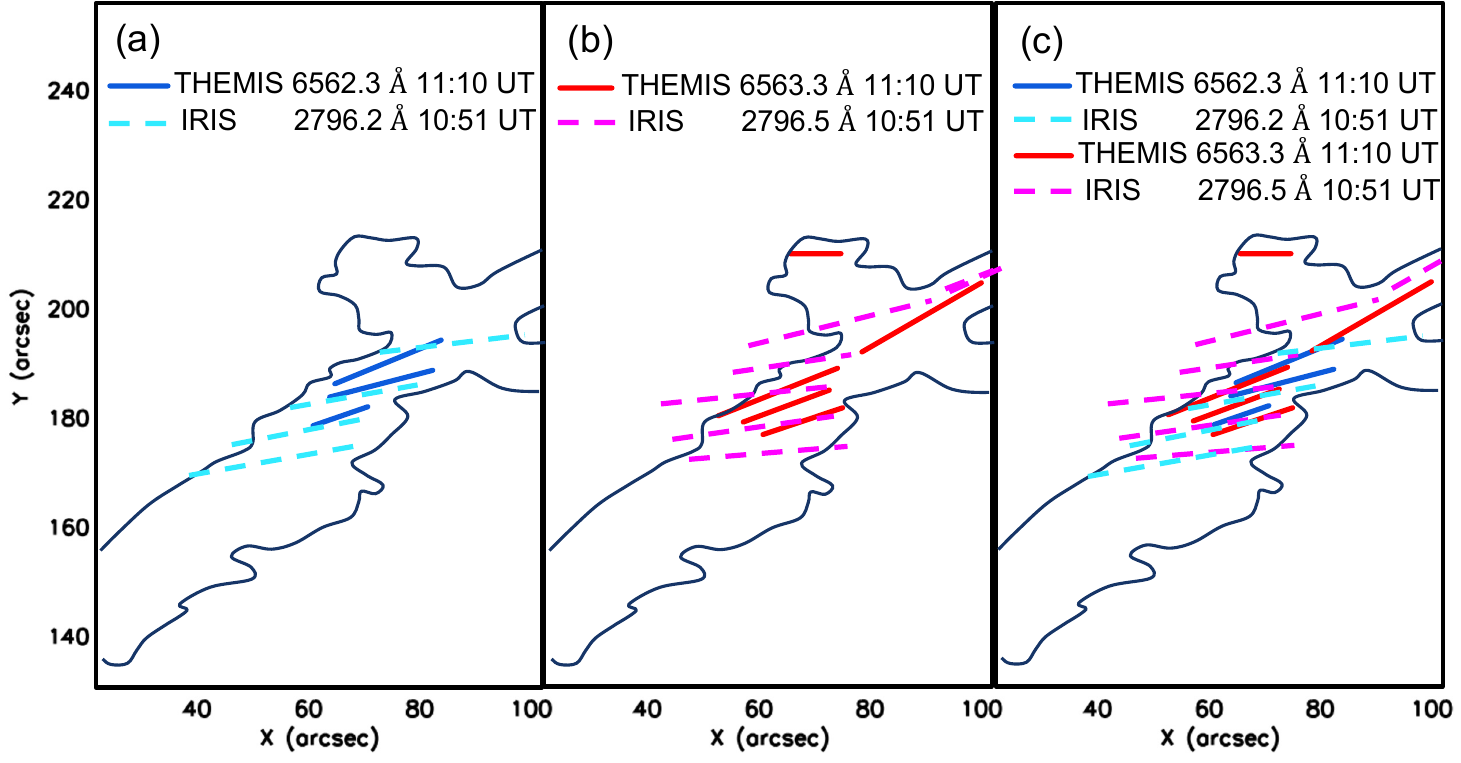}}
\caption{Comparison of H$\alpha$ (continuous lines) and Mg II (dashed lines) high speed components in the blue wing (a),  red wing (b) and both wings (c). These Doppler strands are defined in Figures \ref{map_ha} and \ref{IRIS_maps}.
The FOV of this figure is same as that of white box in Figure \ref{IRIS_maps}).
}
\label{THEMIS_IRIS_compare}
\end{figure}

\begin{figure}    
\centerline{\includegraphics[width=1\textwidth,clip=]{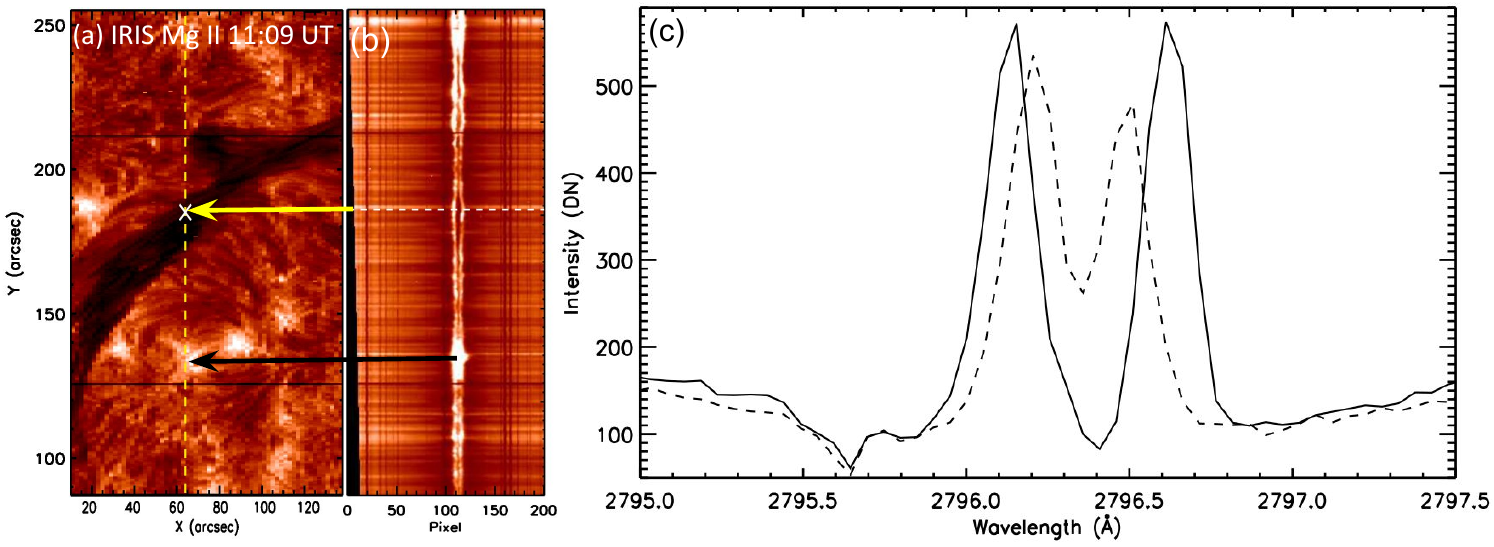}}
\caption{Example of  Mg II k line spectrum and profiles in the filament. Panel (a) presents the intensity reconstructed map derived from the Mg II spectra, panel (b) the IRIS Mg II spectrum of the k line, while panel (c) shows two examples of Mg II k line profiles. The vertical dashed line in panel (a), indicates the spatial location of the spectrum shown in panel (b). In panel (a) the cross in the filament and  in panel (b) the dashed horizontal yellow line mark the location corresponding to the profile with continuous style shown in panel (c).  This profile presents a very broad absorption core, due to the filament absorption, in comparison with the reference profile (dashed line).}
\label{IRIS}
\end{figure}

\subsection{IRIS Maps and Spectra}
\label{subS:IRIS_maps}

Figure \ref{IRIS_maps}
shows the maps obtained  in the two peaks of  the Mg II k line in the reference profile, which represent blue shifted and red shifted Doppler images. As in Figure \ref{map_ha} we have overlaid   with blue and red dashed lines the mostly blue/red shifted strands observed in  the Mg II k Doppler maps (Figure \ref{IRIS_maps}). Here, in between the dark Doppler strands, in the blue$/$red wings, the chromosphere appears, so the optical depth $\tau$ is small for the filament there.  The dark strands are alternately red shifted and blue shifted, indicating that there is counter-streaming.  

There is a good global correspondence for the blue$/$red shift in Mg II and in H$\alpha$, with the northern side of the filament more red shifted than the southern side.
A closer comparison is shown in Figure~\ref{THEMIS_IRIS_compare} where we report the co-aligned dashed lines of Figures \ref{map_ha} and \ref{IRIS_maps}. The Doppler strands observed in H$\alpha$ blue and red wings (continuous lines) are both more aligned with the filament axis than the ones observed in the Mg II k line (dashed lines).  The Mg II k Doppler strands have a direction closer to the chromospheric fibrils observed on both sides of the filament (Figure \ref{map_ha}).  
We conclude that these observations correspond to different heights, with H$\alpha$ showing Doppler strands higher in the filament (close to its top, or spine) and Mg II k line showing Doppler strands at intermediate heights between filament top and chromosphere. In summary, there is a gradient of magnetic shear with height.

An example of spectra in the Mg II line inside the filament is shown in Figure \ref{IRIS}. The filament is visible between y $\approx 170''$ and $\approx 190''$ along the dashed vertical line of panel (a). 
There, the filament width is around 20$''$, so similar as in H$\alpha$. 
The Mg II k spectra in the filament has a dark core due to the absorption of the incident radiation by cool plasma (panel b). In panel (c) the Mg II k line profile is almost symmetric, and it is broader in the filament (continuous line) compared with the reference profile (dashed line). The reference profile is selected from a quiet Sun region, located far from the filament.

The Mg II lines have been deeply studied by \citet{Pereira2013}. They found a good correlation between the simulated plasma velocity at $\tau = 1$ and the Doppler shift measured for the peaks (k2) of the Mg II k line (see their Figure 4). 
Therefore, the displacement of the peaks is a good proxy for the estimation of the plasma velocity along the LOS.

In our case, the displacement of the peaks (k2) corresponds to $\Delta \lambda \approx 0.1$ \AA\ which is equivalent to a velocity of $\pm$ 10 \kms.  The nearly symmetric profile indicates that along the LOS red and blue shifted strands are present. The smoothing of extreme velocity values is expected as a consequence of the limited spatial resolution; then, the deduced velocity is expected to be an underestimation of real plasma velocities.

\subsection{Magnetic Configuration of the Filament}
\label{subS:Magnetic_configuration}

Decades of research have concluded that the dense and cold plasma of quiescent filaments is supported against gravity in almost horizontal magnetic fields (sub-Section \ref{subS:Filament_formation}).  More precisely, with plasma mostly frozen-in the magnetic field (ideal MHD) a stable equilibrium is present in magnetic dips.   The cold plasma is only allowed to move along field lines, and the very small transverse thermal conduction allows the formation of fine structures elongated along field lines. The Doppler strands observed with THEMIS and IRIS are a trace of the magnetic field organization.

The filament strands observed in H$\alpha$ and Mg II line centers have nearly the same orientation (Figure \ref{THEMIS_IRIS}). They are oriented at a small angle with the filament axis. In the line centers the optical depth is large so mainly the filament top is observed. This is consistent with flux rope models where the higher dips are just below the flux rope axis and where the magnetic field is nearly aligned with the flux rope axis \citep[e.g.,][]{Aulanier1998_D}.

As observations are selected further in the spectral line wings the optical depth is
lower, and individual Doppler strands can be better observed. Around the inflection point of the spectral lines, the Doppler strands are darker than the chromosphere at the same wavelength, and the optical depth is large enough that we mostly observe filament plasma in motion. THEMIS Doppler strands have an orientation close to the filament axis (Figures \ref{map_ha} and Figure \ref{THEMIS_IRIS_compare}). In contrast, IRIS velocity threads have an orientation in between the filament axis and the fibrils visible at the chromospheric level on both filament sides (Figure \ref{IRIS_maps}). Thus, these observations trace a sheared magnetic configuration with a magnetic shear small at the chromospheric level and increasing with height as expected in flux rope models \citep[e.g.,][]{Aulanier2003}.

\subsection{Plasma Velocity }
\label{subS:Velocity}

Doppler velocities have been estimated with H$\alpha$ and Mg II k lines (Figures  \ref{map_ha} -- \ref{IRIS_maps}) and the plane of sky velocities with AIA 193 \AA\ filter (Figure \ref{time_slice}). From both observations, one may reconstruct the vector velocity if these observations are probing the same plasma.  Since, the spatial resolution of the instruments as well as the observation techniques are significantly different, one first needs to check if the same plasma structures are observed. In particular, are these observations compatible with our present modeling knowledge of filaments?  As described above, the filament plasma is mostly located in magnetic dips. Since these dips are typically shallow, especially near the filament top (since close to the FR axis), plasma is expected to be able to flow along nearly horizontal field lines.  The  presence of a strong magnetic field (low $\beta$ plasma) implies independent plasma motions in nearby field lines, with a velocity linked to the plasma pressure gradient present independently on each individual field line.

For an horizontal velocity $V$, the east-west component of the velocity is $V \cos \delta$ where $\delta$ is the tilt angle of the strand (inclination to the equatorial plane). This contributes to the Doppler velocity as $V \cos \delta \sin \phi$ where $\phi$ is the longitude from the central meridian. The part of the filament studied is at $\phi \approx 5^0$, so $\sin \phi \approx 0.09$. 
The tilt of H$\alpha$ strands is $\delta \approx 20 \pm 4^0$ and for Mg II k $\delta\approx 9 \pm 4^0$, so the contribution to Doppler velocity is $V \cos \delta \sin \phi \approx 0.08~V$.
Next, the north-south contribution is $V \sin \delta \sin \theta$ where $\theta$ is the latitude of studied part.  With $\theta \approx 18^0$, the contribution to the Doppler velocity is $\approx 0.11~V$ and $\approx 0.05~V$ for H$\alpha$ and Mg II k lines, respectively.
The sum of the contributions leads to Doppler velocities $\approx 0.19~V$ and $\approx 0.13~V$ for H$\alpha$ and Mg II k lines, respectively.   

The plane of sky velocity estimated with the AIA 193 \AA\ filter is close to the horizontal velocity $V$ (since the projection involves only cosines with low value angles). Then, from Figure \ref{time_slice}, $V$ is between $\approx 10$ and 40 \kms . This implies expected Doppler velocities between $\approx 1$ and 8 \kms , so significantly lower than the observed Doppler velocities (of the order of 10 -- 20 \kms ). A first difference between THEMIS, IRIS and AIA observations is the spatial resolution and indeed the fine Doppler strands could be obtained only with the higher spatial resolution of THEMIS and IRIS.  A second major difference is that the AIA observations observed the full EUV continuum absorption profile within the filament while both THEMIS and IRIS provide spectral line profiles that allows the analysis of the spectral wings where Doppler strands could be identified individually, in contrast than in the line core.  Moreover, only low Doppler velocities could be deduced from the spectral line cores.

We conclude that both high spatial and spectral resolution is needed to resolve the Doppler strands.  They correspond only to a small fraction of the filament plasma, as most of the light absorption is done in the spectral line cores. Then, an imager with a broad filter, such as AIA 193 \AA , cannot observe the fraction of plasma moving at larger velocities than the bulk of the filament plasma. Finally, we cannot combine the plane of sky velocity observations of AIA with the Doppler velocities obtained with THEMIS and IRIS.   
It remains that, if we suppose horizontal motions in the filament, the filament spatial location implies horizontal velocities larger than the Doppler velocities by a factor 5 to 8.  A Doppler velocity of 20 \kms\ is already supersonic (the sound speed $\approx 15$ \kms\ in a plasma at a temperature of $\approx 10^4$ K).  Then, this would implies unrealistic strongly supersonic flows (100 to 160 \kms ). 
We conclude that the observed Doppler velocities are probably not due to the field aligned plasma flows in the filament.

Oscillations of the magnetic structure are a plausible explanations for the above observations. Oscillations in filaments have been broadly studied both observationally and theoretically for decades \citep{Arregui2018}.  Here, mainly strands (or threads) oscillations are relevant. The magnetic field line footpoints are excited at small scales by the granule motions, then oscillations perpendicular to the main magnetic field  direction are generated and the frozen-in plasma is forced to move.  Pioneer observational papers found such oscillations in filaments \citep{Thompson1991,Yi1991}. Later these oscillations were identified mostly as kink mode waves \citep[e.g.][]{Lin2003,Lin2009}.
With a velocity amplitude $V_a$ and a period $T$, a sinusoidal oscillation has an amplitude $T \,V_a/(2\pi)$.  With $T \approx 200$ s (strands are oscillating with periods of a few minutes) and $V_a \approx 20$ \kms\ (typical observed Doppler velocity), the wave amplitude is $\approx 1$ Mm, in agreement with the results of \citet{Okamoto2007} in a prominence. This is much smaller than the expected length of a few 100 Mm of the field lines supporting the dense filament.   Then, such  kink waves are only small perturbations of the global magnetic configuration.  If the wave is not fully standing, the evolution of its spatial shape can induce plasma motions parallel to the magnetic field, especially in the dip part which is shallow.  This is another plausible origin of the observed plane of sky flows (Figure \ref{time_slice}).  However, with present THEMIS and IRIS observations we do not have the time cadence to analyze the wave properties, so we rely on previous oscillation studies.

\section{Conclusion and Discussion}
\label{Conclusion}

We study the fine structures present in a quiescent filament observed on the solar disk on 29 September 2023. We analyze the H$\alpha$ and Mg II k spectral lines observed with high spatial and spectral resolutions as achieved with THEMIS and IRIS. We complement these data with AIA and HMI observations to set the previous data within the global context. From the AIA 193 \AA\ filter data we estimate the plane of sky velocity within the filament, and from HMI we deduce a typical magnetic configuration where the filament is above the photospheric inversion line of a bipolar region.

We first build 2D spatial maps in the H$\alpha$ wings (at the inflection point of the H$\alpha$ profile).  Doppler strands of length between 10$''$ and 20$''$ are present in the filament both in the blue and red wings. These blue$/$red strands are interlaced with distance between strands on the order 1$''$ and they form a small angle with the filament axis.  They represent counter-streaming velocities in adjacent Doppler strands.  A similar pattern is observed with Mg II k spectral line, with Doppler strands oriented at a greater angle with the filament axis. In both spectral lines supersonic flows are present (velocities of the order of 20 km s$^{-1}$).  The chromospheric fibrils on both sides of the filament are oriented even further from the filament axis.  We conclude that the magnetic field has an increasing shear with height with, at the filament top, a close alignment with the filament axis and the photospheric inversion line.  

The above results obtained at fixed wavelengths, are confirmed and measured by the analysis of the line spectra. H$\alpha$ has profiles typically broader than the chromospheric reference profiles. Also depending on the spatial location, H$\alpha$ profile is more extended on the blue/red side. Furthermore, several Doppler strands are present in the same H$\alpha$ profile, which is complex with several local absorption features with different Doppler shifts (Doppler strands, Figure \ref{cut_zoom}).  
This result was possible because of the high spatial resolution of the data obtained with THEMIS telescope controlled by the new adaptive optics. 
The same conclusion is reached with the Mg II k spectra. In the filament, Mg II k peaks are typically more separated than for chromospheric spectra. This larger separation is a signature of absorbing red and blue shifted plasma. However, this conclusion is in contradiction with previous prominence studies where coherent flows were found in larger Doppler strands than fine intensity structures.  For example, \citet{Schmieder2010} analyzed the fine intensity structures observed in H$\alpha$ by Hinode/SOT and by the Meudon solar tower, obtaining Doppler shifts via bisector analysis of MSDP data. Finally, the result difference between this study and the present one are due to the difference of the spatial and spectral resolutions of the instruments used.

We interpret all the above results within a locally dipped magnetic configuration. The plane of sky velocities deduced from the AIA 193 \AA\ filter data represent global motions along the filament.  These velocities cannot be combined with the observed Doppler velocities, first because of the spatial resolution difference between AIA and THEMIS and IRIS, and second because only a small fraction of the plasma is Doppler shifted while AIA observes the whole filament plasma.

Such dynamics could not be detected in  previous observations made, for example,  with the MSDP instrument with only a few points in the spectral line, and the wavelength domain extended only to $\pm$ 0.6 \AA\ \citep[e.g.][]{Schmieder1985}.
This is a spectral domain for testing the dynamics in prominences that have relatively low velocity ($<<$ 20 \kms). Moreover, the Doppler velocity is derived with the bisector method which provides only a ``mean" velocity between the blue and red shifted wings.  Therefore, the measured velocities were small within the filament.
The spatial resolution was also not enough to detect fluctuations of velocities in structures around 1$''$ to 6$''$ wide.  In contrast, the analysis of Doppler maps in the far wings of H$\alpha$ with filters allow the detection of counter-streaming velocities \citep{Zirker1998_c,Lin2003,Lin2009, Schmieder2008f}. We should also take into account that this method makes the assumption that the filament is optically thick and that the chromosphere below has no impact in the H$\alpha$ profiles.

Here, with THEMIS and IRIS high-resolution spectra we have made a step forward to detect supersonic flows in sub-arcsecond Doppler strands.
Since the observed filament is located relatively close to the disk center, these Doppler velocities would imply horizontal velocities about 10 times larger than the sound speed if they were interpreted as counter-streams present along the nearly horizontal magnetic field in the filament.  Rather, these Doppler velocities are expected to be due to kink oscillations of the strands with modest amplitudes (about 1 Mm).  Each strand looks to have a coherent motion, different from its neighbors, so the presence of mixed blue-red shifted strands.  However, the studied data do not have the temporal resolution to analyze such wave patterns.  
The speeds found with the Doppler shifts imply supersonic flows if they are interpreted as magnetic field aligned flows. These horizontal flows could be interpreted as transient  high-speed motions in elongated threads by thermal non equilibrium models \citep{Karpen2006}. However, no such high longitudinal flows were detected in our present observations with AIA. This may be due to its limited spatial resolution.

These kinds of observations should be extended with higher spatial and spectral resolution with the Swedish Solar Telescope (SST), and Daniel K. Inouye Solar Telescope (DKIST) telescopes.  It will be important to obtain time series to analyze the possible oscillations in order to extend previous results  \citep[e.g.][]{Thompson1991, Arregui2018}.  
Performing non-linear force-free extrapolations and MHD simulations would complement our understanding of the magnetic field configuration, setting the observed velocities in the framework of arcade or flux rope model for filament support \citep{Guo2010,Guo2022}.


\begin{acks}
We would like to thank the editor Prof. Arregui for the smooth and helpful editorial process. We would like to acknowledge the reviewer for constructive and insightful comments, which have significantly improved this work. This work is based on ground-based observations obtained by the THEMIS telescope in Tenerife in the Canary Islands, operated by Bernard Gelly, Richard Douet, and Didier Laforgue during a multi-wavelength campaign with IRIS (IHOP444 - PIs Nicolas Labrosse - Brigitte Schmieder).
The  H$\alpha$ spectroheliograph was provided by BASS2000.
obspm.fr. AIA data are courtesy of NASA/SDO and the AIA, EVE, and HMI science teams. IRIS is a NASA small explorer mission developed and operated by LMSAL with mission operations executed at NASA Ames Research Center and major contributions to down-link communications funded by ESA and the Norwegian Space Centre.
\end{acks}

\begin{authorcontribution}
GK did the data analysis and wrote the draft of the paper. BS, PDs, RC and RJ wrote substantial parts of the manuscript and contributed to the interpretation. All the authors did a careful proofreading of the text and references.  BS has obtained this observation during the observational compaign at THEMIS along with the  project IHOP444 team  (PIs N.Labrosse and B.Schmieder).
\end{authorcontribution}
\medskip
\noindent {\scriptsize {\bf Funding} GK acknowledges the support form DST/INSPIRE. BS was supported by the SOLARNET EU program under the project ``Magnetic field structure of prominences, solar tornadoes and spicules". R.C. acknowledges the support received from the DST/SERB under project number EEQ/2023/000214.
BS thanks the  LUNEXLunar Explorers Society, SBIC,  to support the presentation of this work at the IRIS meeting in UK.}
\medskip

\noindent {\scriptsize {\bf Data Availability} The datasets used in the present study are available at {\url {http://jsoc.stanford.edu/}}, {\url {https://iris.lmsal.com/search/}}, {\url{https://bass2000.obspm.fr/home.php}}. }

\begin{conflict}
 The authors declare that they have no conflicts of interest.
\end{conflict}

\bibliographystyle{spr-mp-sola}
\bibliography{references}  

     

\section{Appendix}
\label{Appendix}

\begin{figure}[h]    
\centerline{\includegraphics[width=0.75\textwidth,clip=]{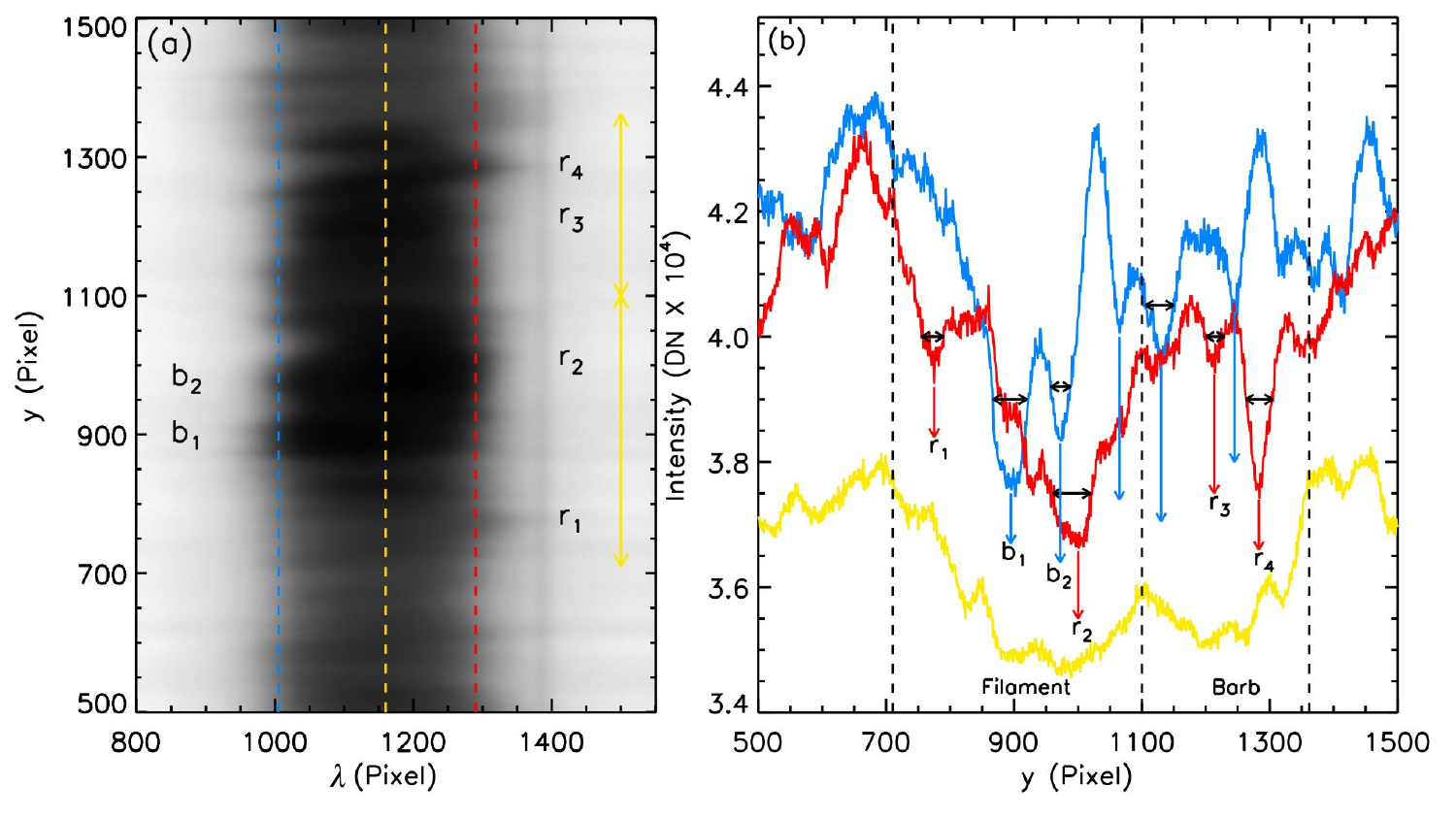}
}
\caption{H$\alpha$ spectrum ($\lambda$, y) at 11:05 UT for the spectra 107 (panel a) and cuts along the slit at three wavelengths (panel b) showing the velocity shift in the filament, the barb and the chromosphere.   
In panel (a), the dashed vertical lines with blue, yellow, and red colors correspond to velocity of -20, 0, and 20 \kms, respectively. These lines show the cut locations used for the curves with same colors of panel (b). The positions of the filament and the barb are indicated by two yellow double vertical arrows in panel (a) and by vertical dashed black lines in panel (b).
In panel (b) the high Doppler strands in the blue wing (b$_1$, b$_2$, ...) and in the red wing (r$_1$, r$_2$...) are indicated by double black horizontal arrows.  The vertical arrows mark their Doppler shifts with typically alternating blue and red arrows.}
\label{cut_1105_Barb}
\end{figure}

\begin{figure}[h]   
\centering{\includegraphics[width=0.8\textwidth,clip=]{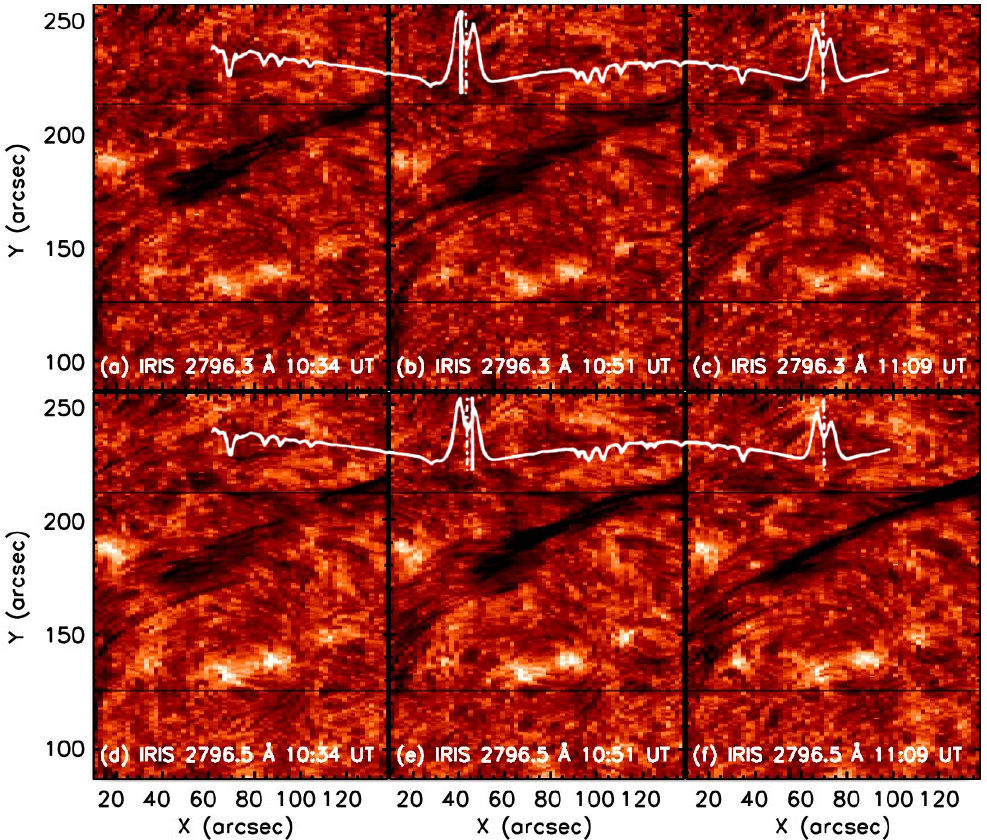}}
\caption{IRIS reconstructed maps at the peaks of Mg II k lines  (V= $\pm$ 10 \kms) obtained during 3 rasters taken at (a, d) 10:34 UT, (b, e) 10:51 UT and (c, f) 11:09 UT.}
\label{IRIS_2}
\end{figure}

In this appendix we show other times of THEMIS and IRIS observations.  
Figure \ref{cut_1105_Barb} complements Figure \ref{cut_zoom} and Figure \ref{IRIS_2} complements Figure \ref{IRIS_maps}.

\vspace*{-0.50cm}

\end{document}